\documentclass[aps,prl,twocolumn,showpacs,superscriptaddress,groupedaddress,longbibliography,showpacs]{revtex4-1}
\usepackage{graphicx}
\usepackage{newtxtext,newtxmath}
\usepackage{enumerate}
\usepackage{bm}
\usepackage{dcolumn}
\usepackage{geometry}
\begin{document}

\title{Lattice deformation on flat-band modulation
in 3D Hopf-linked carbon allotrope: Hopfene}

\author{Isao Tomita}
\email{i.tomita@soton.ac.uk}
\affiliation{Sustainable Electronic Technologies,
Electronics and Computer Science,
Faculty of Physical Sciences and Engineering,
University of Southampton, SO17 1BJ, UK}
\affiliation{Department of Electrical and Computer Engineering,
National Institute of Technology, Gifu College, Gifu 501-0495, Japan}
\author{Shinichi Saito}
\affiliation{Sustainable Electronic Technologies,
Electronics and Computer Science,
Faculty of Physical Sciences and Engineering,
University of Southampton, SO17 1BJ, UK}


\begin{abstract}
Flat bands form in a 3D Hopf-linked graphene crystal
or a 3D carbon allotrope named Hopfene, which
qualitatively differ from bands of only graphenes.
This paper discusses carbon-hexagon deformation
on the level shift of a flat band via density-functional-theoretical
(DFT) analysis to set the flat-band level
to the Fermi level, viz., to utilize its large density of states
for magnetic- and electronic-property researches.
Tight-binding (TB) analysis is also performed for a comparison
with the DFT analysis; here, a qualitative agreement
between TB and DFT bands is obtained.  The DFT analysis
shows an almost linear flat-band level shift
to the lattice-deformation rate, where electron-interaction
effects are included within the Kohn-Sham method.
To tune the flat-band level so that it fits the Fermi level,
a double-hetero-like structure is also proposed
as a way of hexagon-deformation control.
\end{abstract}

\maketitle

Carbon allotropes that have been found for the past
three decades, such as fullerenes \cite{Kroto},
nanotubes \cite{Iijima1}, and graphenes \cite{Nov1},
have uncovered that {\it morphology} (even though
the material is the same) greatly changes
the fundamental mechanical, electrical, and optical
properties of the material \cite{Iijima2,Ando,Geim,Tanaka}.
These opened very useful device and material
applications in nano-electronics and material science,
and also developed basic physics researches related to
Tomonaga-Luttinger liquid \cite{TL}, superconductivity
\cite{Super}, quantum Hall effect \cite{QH1},
and Kosterlitz-Thouless transition \cite{KT}.
The ground-breaking discovery of fullerenes and
graphenes, linked to those advanced fundamental
and applied researches, won Nobel prizes in 1996 and
2010, respectively.

\begin{figure}[htbp]
\begin{center}
\includegraphics[width=80 mm]{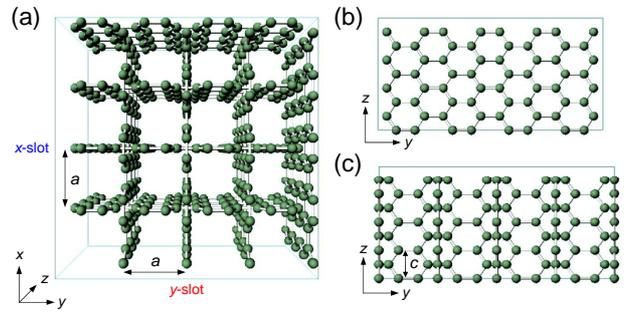}
\end{center}
\caption{(a) Hopfene made of horizontally- and
vertically-combined graphene sheets where
the intersections are comprised of Hopf-links;
this Hopfene has an empty slot between neighboring
graphene sheets \cite{S+T_1,T+S_1}. (b) Graphene sheet
before vertical graphene sheets are inserted.
(c) Graphene sheet after they are inserted; this insertion
swells carbon hexagons in the $y$ (and $x$) direction.}
\label{fig1}
\end{figure}

Morphology, including topological one, has now been
attracting a great deal of attention not only in physics
dealing with the above allotropes and related researches
but also in chemistry treating artificially-designed molecules
that can be used as molecular machines \cite{Chem}; the latter
also won Nobel prize in 2016.  As for carbon structures,
we have recently developed other types of allotropes
from a topological point of view \cite{S+T_1}.  In a crystal-type
topological allotrope among them, named {\it Hopfene}
(see Fig.~\ref{fig1}(a)), we have revealed some unusual
electronic properties by band-structure analyses \cite{T+S_1},
and also have found striking 3D-Dirac/Weyl-fermionic features
\cite{S+T_2}.  Here, the terminology {\it Hopfene}
originates from a crystal with {\it Hopf}-linked
intersections, named after topologist Heinz Hopf
who studied these links greatly.

Until now we have shown structural sustainability
of Hopfenes via semi-empirical molecular-orbital
\cite{S+T_1} and density-fractional-theoretical (DFT)
\cite{T+S_1,DFT1,DFT2} methods (although the graphene-sheet
insertion parallel to the $x$-$z$ plane
(the $y$-$z$ plane) swells the carbon hexagons
in the $y$ direction ($x$ direction), as depicted
in Figs.~\ref{fig1}(b)(c)).
In this paper, we examine this hexagon swelling
on the band structure, particularly on flat bands,
via DFT analysis.  Before doing this by heavy numerical
computations, we carry out simple
tight-binding(TB)-Hamiltonian analysis, which is frequently
employed in analyzing graphene band-structures.
Later in this paper, those DFT- and TB-analysis results
will be compared with each other, particularly
to determine transfer-energy parameters
for Hopf-links in the TB model.

\begin{figure}[htbp]
\begin{center}
\includegraphics[width=80 mm]{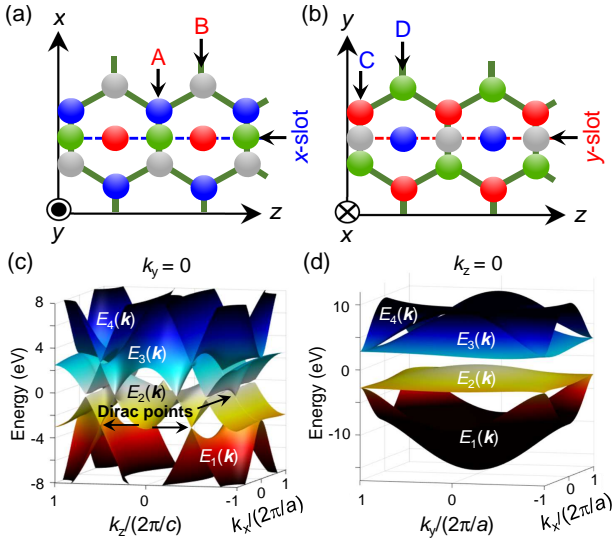}
\end{center}
\caption{(a) Graphene sheet parallel to the $x$-$z$ plane
that defines A- and B-sublattices, including Hopf-links,
with $x$-slot insertion.
(b) Graphene sheet parallel to the $y$-$z$ plane
that defines C- and D-sublattices, including Hopf-links,
with $y$-slot insertion.
(c) Band structure in the $k_x$, $k_z$ directions,
obtained by diagonalizing Eq.~(\ref{eq1})
with $t_{\rm G}=2.8$ eV and $t_{\rm H} = 1.0$ eV
for $k_y =0$.  Dirac points reflecting those of
graphene sheets are observed.
(d) Band structure obtained with the same $t_{\rm G}$
and $t_{\rm H}$ in the $k_x$, $k_y$ directions
for $k_z =0$, which exhibits a flat band.}
\label{fig2}
\end{figure}

A difference in Hamiltonian between Hopfenes and graphenes
is that since the Hopfenes have Hopf-linked intersections
(where graphenes meet), the total Hamiltonian has an extra
component ${\cal H}_{\rm H}$ concerning electron transfer
via the Hopf-links in addition to ordinary graphene transfer
Hamiltonians ${\cal H}_{\rm G}(h_{\rm AB})$ and
${\cal H}_{\rm G}(h_{\rm CD})$ defined
in A- and B-sublattices on the $x$-$y$ plane
and C- and D-sublattices on the $y$-$z$ plane
(see Figs.~\ref{fig2}(a)(b)). The presence of
${\cal H}_{\rm H}$ affects Dirac-point symmetries
that graphenes originally have \cite{S+T_2}.
Note that there is a difference in theory-applicable range
between TB and DFT analyses, where TB analysis
can deal with perfect hexagons and DFT analysis can
{\it also} deal with deformed hexagons, which is
out of reach of TB approach; yet TB approach is sometimes
{\it still} effective in extracting important basic
features of electronic bands, which will be
shown below.

The total TB Hamiltonian \cite{S+T_2} is given by
\begin{equation}
\hat{H} = \sum_{{\bf k}, \sigma}
\sum_{\mu, \nu =1}^2
{\hat{\psi}}_{{\bf k} \sigma}^{\dagger (\mu)} \,
{\cal H}^{\rm total}_{\mu \nu} \, 
{\hat{\psi}}_{{\bf k} \sigma}^{(\nu)},
\label{eq1}
\end{equation}
where ${\hat{\psi}}_{{\bf k} \sigma}^{(1)}
=(a_{{\bf k} \sigma}, b_{{\bf k} \sigma})$
and ${\hat{\psi}}_{{\bf k} \sigma}^{(2)}
=(c_{{\bf k} \sigma}, d_{{\bf k} \sigma})$
are annihilation operators for electrons
of momentum $\bf k$ (in units of $(2 \pi/a, 2 \pi/a,
2 \pi/c)$) and spin $\sigma$ in A- and B-sublattices
and C- and D-sublattices, respectively,
which are basically equivalent, except for plane directions;
their adjoints ${\hat{\psi}}_{{\bf k} \sigma}^{\dagger (1)}$
and ${\hat{\psi}}_{{\bf k} \sigma}^{\dagger (2)}$
are electron-creation operators; and 
${\cal H}^{\rm total}_{\mu \nu}$ are defined as
${\cal H}^{\rm total}_{11}={\cal H}_{\rm G}(h_{\rm AB})$,
${\cal H}^{\rm total}_{22}={\cal H}_{\rm G}(h_{\rm CD})$,
${\cal H}^{\rm total}_{12}={\cal H}_{\rm H}$, and
${\cal H}^{\rm total}_{21}={\cal H}_{\rm H}^{\dagger}$
(these are all $2 \times 2$ matrices).
For a (2,2)-Hopfene,
$[{\cal H}_{\rm G}(h_i)]_{11} = [{\cal H}_{\rm G}(h_i)]_{22} = 0$,
$[{\cal H}_{\rm G}(h_i)]_{12}
= [{\cal H}_{\rm G}(h_i)]_{21}^{*} = h_i$
($i$ = AB, CD) with $h_{\rm AB} = 
 - t_{\rm G} ( e^{{i k_x a}/{3}}
+ 2 e^{-{i k_x a}/{6}} \cos ( {k_z c}/{2} ))$,
$h_{\rm CD} = 
 - t_{\rm G} ( e^{{i k_y a}/{3}}
+ 2 e^{-{i k_y a}/{6}} \cos ( {k_z c}/{2} ))$,
and ${\cal H}_{\rm H} = - t_{\rm H}(
\phi_{-{k_x}/{6}}^{\dagger} \phi_{{k_y}/{3}}
+\phi_{{k_x}/{3}}^{\dagger} \phi_{-{k_y}/{6}}
+2 \cos ({k_z c}/{2})
\phi_{-{k_x}/{6}}^{\dagger} \phi_{-{k_y}/{6}})$
with $\phi_k = ( e^{i k a}, e^{-i k a})$;
here $t_{\rm G}$ and $t_{\rm H}$ are
transfer-energy parameters \cite{S+T_2}.
For $t_{\rm G}$, there is a good reference
\cite{Ando,RMP}, but for $t_{\rm H}$, there is
no such one (and we will obtain this
in comparison with the DFT analysis).

If holes are dealt with (when the Fermi level
is below a Dirac point), the annihilation and
creation operators given above should be
switched to those for holes.  Moreover,
the spin-orbit interaction, which predicts
the appearance of 'spin-Hall' effect, is
not included in Eq.~(\ref{eq1}), since
we here discuss no such phenomenon as current
induction by this effect (and this effect is actually
very small in light materials, e.g., carbons,
when compared with that in heavy materials,
e.g., HgTe-quantum wells).

By diagonalizing Eq.~(\ref{eq1})
\cite{Ando,AS} in a numerical manner,
we obtain energy-momentum dispersion
$E_i ({\bf k})$ with band index $i$,
as depicted in Fig.~\ref{fig2}(c) in the $k_x$, $k_z$
directions with $k_y =0$, where $t_{\rm G}$ is
set at 2.8 eV \cite{Ando,RMP} and $t_{\rm H}$ is
tentatively set at 1.0 eV.  Here we can see
some Dirac points reflecting those of graphenes;
but the fine detail is different from graphenes:
the equivalence of the original six Dirac points
is violated by the presence of the Hopf-links,
and partially degeneracy-resolved
Dirac points appear \cite{S+T_2}.

A more remarkable difference from the graphenes
is the presence of a big flat band, as illustrated
in Fig.~\ref{fig2}(d) in the $k_x$, $k_y$ directions
with $k_z =0$; much more flat-band appearances
are recognized in a (2,2)-Hopfene with an empty slot
between two adjacent graphene sheets
(with atomic symmetry P4$_2$)
than in a (1,1)-Hopfene with a similar
structure but a smaller (or minimum, finite)
sheet-spacing (with P4$_2$bc) \cite{T+S_1}.

The flat bands, which provide high density of states (DOS),
are useful for improving magnetic and electric properties
of crystals, e.g., ferromagnetism \cite{FB1} and
superconductivity \cite{FBS}, where the flat bands much
enhance the electron-electron interaction while suppressing
one-particle degree of freedom (which is beneficial
for such improvement).  To make use of those high-DOS
flat bands, the level of a flat band should fit
the Fermi level.  But this is not always satisfied,
which actually depends on the band structure and
electron filling in a crystal; the flat-band-level
tuning is thus necessary.  We perform this by carbon-hexagon
deformation in the Hopfene.

In addition to Hopfenes, there have been some 3D carbon
allotropes proposed so far that show flat bands
\cite{Koshi,Zhong,Hua}.  For effective use of flat-band levels,
Fermi-level tuning by means of doping or intercalation
is used for the level-fitting; but too much doping
for a large Fermi-level shift completely spoils
the band structure. Instead, we here describe a flat-band shift
by hexagon-deformation while averting this band-structure
spoiling due to too much doping, and perform precise
level-tuning while keeping the flatness of the band;
this enables large level-tuning
($\sim {\rm eV}$).  This type of research has not fully
been performed yet in other 3D carbon allotropes.

\begin{figure}[htbp]
\begin{center}
\includegraphics[width=80 mm]{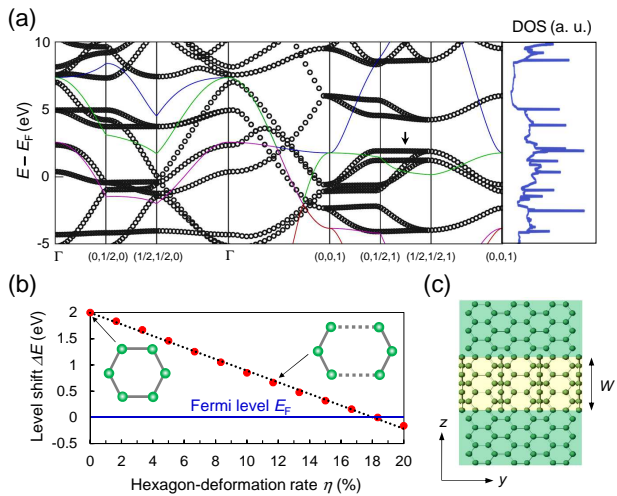}
\end{center}
\caption{(a) Band structure of a (2,2)-Hopfene. The black circles
come from DFT computations, and thin colored curves
come from TB calculations via a fitting of $t_{\rm H}$.
(b) Level shift $\Delta E$ (eV)
of a flat band indicated with the black arrow in (a)
depending on the hexagon-deformation rate $\eta$ (\%),
where $\Delta E$ is measured from the Fermi level $E_{\rm F}$.
At $\eta = 18.2$ \%, the flat band matches $E_{\rm F}$;
such a large c-c-bond extension is known to be allowed
for graphenes \cite{Sorella}.  (c) Double-hetero-like
structure for hexagon-deformation control in the Hopfene (yellow),
where the Hopfene is sandwiched between graphene layers
(green).  Adjusting the size of $W$ controls $\eta$.}
\label{fig3}
\end{figure}

We find via DFT analysis that the sheet insertion
(which swells the carbon hexagons in the $x$, $y$ directions)
can shift the flat-band level, where we start from perfect hexagons
in a (2,2)-Hopfene, and see an effect of gradual hexagon-deformation.
Band structures (Kohn-Sham-eigenvalue distributions)
of a (2,2)-Hopfene with perfect hexagons are plotted
in Fig.~\ref{fig3}(a) for TB (colored curves)
and DFT results (black circles); their comparison in shape
gives qualitatively good agreement (but no exact match),
which gives $t_{\rm H}= 1.5$ eV (with a standard deviation
of 0.5 eV).

Next, we show the level shift of a flat band,
indicated by the black arrow in Fig.~\ref{fig3}(a),
by deforming the perfect hexagon, starting from
the hexagon-deformation rate $\eta = 0$ (\%).
As seen in Fig.~\ref{fig3}(b), the level goes down
almost linearly as $\eta$ (\%) increases.
In Fig.~\ref{fig3}(b), the gradient of
the energy shift $\Delta E$ (eV) vs $\eta$ (\%)
is $- 0.111$ (eV/\%), where $\eta=100 \, \%
\times (a/a_0 - 1)$ with a swelled c-c-bond length $a$
in the $x$, $y$ directions and the original c-c-bond length
$a_0$ (= 1.42 \AA); this means that a 1-\% bond extension
moves the flat-band level down by 111 meV.
This level-shift rate does not so much differ from
that (for holes) produced by strain-induced band-deformation
in semiconductors \cite{Compound}.  The level shift
for electrons in the semiconductors is small compared
with that for holes, particularly for heavy holes,
but the electron level shift in our Hopfene is
comparable to that heavy-hole level shift, which may
be due to dealing with 'heavy' electrons
at the flat band produced by large band-deformation.

So far the linear dependence of $\Delta E$ on $\eta$
has not yet been explained well.
In addition to the strain effect,
there could be interaction effects between electrons
at the c-atoms; more specifically, the c-c-bond swelling
could decrease the repulsive-interaction energy
for electrons localized at the c-atoms in the bonds.
Also, we need to take into account other interaction effects,
since there are $\pi$-electrons distributed extensively
in the crystal \cite{T+S_1}, interacting with each other,
and interacting with the localized electrons at the c-atoms
and also electrons around $\sigma$-bonds.
These complicated interaction effects could be associated
with the relation between $\Delta E$ and $\eta$.

As seen in Fig.~\ref{fig3}(b), at $\eta =18.2$ (\%)
the flat-band level comes across the Fermi level $E_{\rm F}$.
Since $\eta \approx 20$ \% is kept in natural Hopfenes,
those two levels are a bit separated, actually.
To control the $\eta$-change and stop it
at an appropriate $\eta$, a pressure application
is helpful; but a high pressure on the order of GPa
will be necessary.  This size of pressure is possible,
and a super compressor that generates $\sim 100$ GPa
has actually been developed (to observe superconductivity)
\cite{GPa}, but is costly.

We here propose another method that uses
a double-hetero(DH)-like structure, as given
in Fig.~\ref{fig3}(c), which is a similar structure
employed for semiconductor strain-band-engineering.
With this structure, when the width $W$ is controlled,
controlling the hexagon deformation in the $y$ direction
is possible.  If $W$ is small, or comparable to the Hopfene
unit-cell size, the Hopfene hexagons are almost undeformed,
i.e., almost the same as those of graphenes placed
at both sides of the Hopfene.  As $W$ increases,
the deformation becomes large.  Detailed calculations
for appropriate $W$ will be necessary for those levels
to fit exactly.  Here we should note that in our proposed
DH-like structure, deformation controls both in the $x$, $y$
directions are not performed simultaneously;
in Fig.~\ref{fig3}(c), it controls only the flat-band level
in the $y$ direction.  Controlling the level in the $x$ direction
is possible by switching $y$ to $x$ in directions
(but with no simultaneous controls in the $x$, $y$ directions).
This limits an available $k$-space in flat-band modulation,
but does not destroy a level-shifting way
reported in this paper.

In summary, we have examined the band structures of
Hopf-linked graphene crystal named Hopfene
via TB and DFT analyses.  We have seen that
they agree qualitatively well with each other
via TB-parameter fitting.  The hexagon deformation
on a flat-band-level shift in a (2,2)-Hopfene
has been analyzed by the DFT method, which cannot be
easily made by the TB method including free parameters.
That deformation enables fitting the flat band
to the Fermi level (where its large DOS will be used
for magnetic and electric researches \cite{FB1,FBS}).
With an increase in the hexagon-deformation rate $\eta$,
we have observed a linear lowering of the flat-band level,
where the flat band meets the Fermi level at a certain $\eta$.
We also have put forward an idea using a DH-like structure
to control $\eta$ and the flat-band level, without
applying a very high pressure.

\section*{Acknowledgment}
This work is supported by EPSRC Manufacturing Fellowship
(EP/M008975/1). I.T. would like to thank ECS at University
of Southampton, where the idea of the present research
was come up with. (The data of the paper can be obtained
from the University of Southampton ePrint research repository:
https://doi.org/10.5258/SOTON/D0953.)  We would also like
to thank JAIST for its hospitalities during our stay
at the Center for Single Nanoscale Innovative Devices.
Thanks are also due to Prof. H. Mizuta, Dr. M. Muruganathan,
Prof. Y. Oshima, Prof. S. Matsui, Prof. S. Ogawa,
Prof. S. Kurihara, and Prof. H. N. Rutt
for stimulating and fruitful discussions.

\end{document}